\documentclass[aps,prl,twocolumn,groupedaddress]{revtex4-1}

\usepackage{epsf}
\usepackage{graphicx}
\usepackage{amssymb}

\usepackage{color}

\begin{document}

\title{Manifestation of New Interference Effects in  Superconductor/Ferromagnet
Spin Valve}

\author{P. V. Leksin, N. N. Garif'yanov, I. A. Garifullin}
\email{ilgiz\_garifullin@yahoo.com} \affiliation{Zavoisky Physical-Technical Institute, Kazan Scientific Center of Russian Academy of Sciences,
420029 Kazan, Russia}
\author{J. Schumann, V. Kataev, O. G. Schmidt, B. B\"{u}chner}
\affiliation{Leibniz Institute for Solid State and Materials
Research IFW Dresden, D-01171 Dresden, Germany}

\date{\today}

\begin{abstract}
Superconductor/ferromagnet (S/F) spin valve effect theories  based on the S/F proximity phenomenon assume that the superconducting transition
temperature $T_c$ of F1/F2/S or F1/S/F2 trilayers for parallel magnetizations of the F1- and F2-layers ($T_c^{P}$) are smaller than for the
antiparallel orientations ($T_c^{AP}$). Here, we report for CoO$_x$/Fe1/Cu/Fe2/In multilayeres with varying Fe2-layer thickness the sign-changing
oscillating behavior of the spin valve effect $\Delta T_c=T_c^{AP}-T_c^P$. We observe the full direct effect with $T_c^{AP}>T_c^P$ for Fe2-layer
thickness $d_{Fe2}<$1\,nm and the full inverse ($T_c^{AP}<T_c^P$) effect for $d_{Fe2}\geq$1\,nm. Interference of Cooper pair wave functions reflected
from both surfaces of the Fe2-layer appear as the most probable reason for the observed behavior of $\Delta T_c$.



\end{abstract}

\maketitle

More than 40 years  ago Larkin and Ovchinnikov \cite{Larkin}  and Fulde and Ferrel \cite{Fulde} theoretically demonstrated that a strong
spin-exchange field acting on electron pairs in a superconductor may yield superconductivity with a nonuniform order parameter. In a layered
superconductor/ferromagnet (S/F) thin film heterostructure such an exchange field gives rise to a damped oscillating behavior of the Cooper pair
wave function which penetrates from a superconductor into a ferromagnet because of the non-zero momentum of the Cooper pairs in the F-layer. In the
"dirty" limit the characteristic depth of the decay of the pairing function in the F-layer $\xi_m=(4\hbar D_m/I)^{1/2}$ is determined by the the
diffusion coefficient $D_m$ and the exchange splitting $I$ of the conduction band in the F-layer \cite{Radovic}. For pure Fe the value of $\xi_m$ is
less than 1\,nm (see, e.g., \cite{Lazar}). The oscillating behavior of the pairing function in the F-layer results in a number of new experimentally
observed effects: the spatial oscillation of the electron density of states \cite{Ryazanov2001,Kontos2002}, a nonmonotonic dependence of the
superconducting (SC) transition temperature $T_c$ of F/S/F trilayers on the F-layer thickness \cite{Lazar,Garifullin2002} and the realization of the
Josephson $\pi$-junctions in S/F/S systems \cite{Ryazanov2004} (for a review see, e.g., \cite{Buzdin_rev}).

The so-called spin valve effect, a complete on/off switching  of
the SC current flowing in complex S/F multilayers, gives another
example of the fascinating interplay between magnetism and
superconductivity. Its theoretical concept developed in Ref.
\cite{Oh,Tagirov,Buzdin} is based on the S/F proximity phenomenon
and relies on the idea to control the pair breaking and
correspondingly the $T_c$ by manipulating the mutual orientation
of magnetizations of two F-layers.  This is because the mean exchange field
from two F-layers acting on the Cooper pairs
is smaller for antiparallel (AP) orientation of magnetizations
of these F-layers compared to the parallel (P) case. Theories
\cite{Oh,Tagirov,Buzdin} predict that the
AP orientation is always more favorable for
superconductivity than the P one so that $T_c^{AP}$
should be always larger than $T_c^{P}$. However, in case of the
oscillating behavior of the Cooper pair wave function in the
F-layer possible interference effects in the spin valve cannot be
ignored. Especially this concerns the spin valve design F1/F2/S
proposed by Oh et al. \cite{Oh} whose functionality should
obviously critically depend on the interference at the F2/S
interface of the pair wave function reflected from both surfaces
of the F2-layer.

Recently we experimentally realized a full switching effect  for the SC current by changing the mutual orientation of the F-layers' magnetizations in
the system CoO$_x$/Fe1/Cu/Fe2/In \cite{Leksin}, in which the positive difference $\Delta T_c=T_c^{AP}-T_c^P$ quickly decreases from 19 to 12\,mK with
increasing the thickness $d_{Fe2}$ from 0.5 to 0.6\,nm and finally vanishes for the sample with $d_{Fe2}$=2.6 nm. In this system the cobalt
oxide antiferromagnetic layer plays a role of the bias layer which pins the magnetization of the Fe1-layer and Cu is a normal metallic layer which
decouples the magnetizations of the Fe1- and Fe2-layers.

The objective of the present work is the search of a manifestation of the quantum interference effects in an F1/F2/S spin valve system. For that we
have investigated the dependence of $\Delta T_c$ on the thickness $d_{Fe2}$ of the intermediate Fe2-layer on an extensive set of spin valve samples
CoO$_x$(4nm)/Fe1(3 nm)/Cu(4 nm)/Fe2($d_{Fe2}$)/In(230 nm) with the value of $d_{Fe2}$ lying in the range between 0.4 and 5.2\,nm. We obtained clear
experimental evidence for the oscillating behavior of the spin valve effect, i.e. the sign change of $\Delta T_c$ as a function of $d_{Fe2}$, which
we attribute to the interference effect of the SC pairing function occurring in the F2-layer.

We used the same sample preparation method, experimental set ups and protocols of magnetic and transport measurements as in our previous work (see
\cite{Leksin}).

The residual resistivity ratio {\it RRR}=R(300 K)/R(4 K)
(the ratio of the electrical resistivity at 300 K to its value
at 4 K) for all studied samples
is similarly good evidencing a high purity of the deposited In
layers. It lies in the interval 35 $\leq RRR \leq$45 corresponding
to the range of the SC coherence lengths 150 nm
$\leq$ $\xi_s $ $\leq$ 170 nm. For this set of samples the ratio
between the In film thickness $d_s$ and $\xi_s$ is in the range
1.3$\leq$ ${d_s}/{\xi_s}$ $\leq$ 1.5. This clearly shows that the
dominant part of the In layer is involved into the proximity
effect making the $T_c$ particularly sensitive to the
ferromagnetic Fe-layers. Decreasing the In layer thickness down to
$d_s \sim$ 180 nm ($d_s /\xi_s \sim 1.1$) revealed that the
superconductivity in this case is strongly suppressed yielding
$T_c$ below 1.4\,K.

Experimentally we focussed on the determination of both the
hysteresis magnetization behavior and the current in-plane
transport measurements, enabling a correlation between both
properties. To determine the magnetic field range where AP and P
states can be achieved the in-plane magnetic hysteresis loops of
samples in the direction of the magnetic field along the easy axis
were measured by a
\begin{figure}[t]
\centering{\includegraphics[width=0.4\columnwidth,angle=-90,clip]{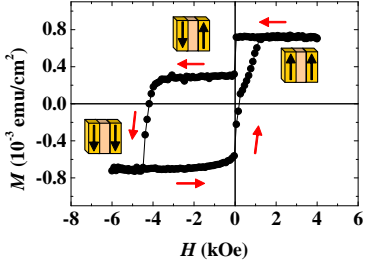}}
\caption{(Color online) The magnetic hysteresis loop for the
sample CoO$_x$/Fe1(2.9 nm)/Cu(4 nm)/Fe2(1.3 nm)/In(230 nm).}
\end{figure}
SQUID magnetometer. Fig. 1 shows the magnetic hysteresis loop  for
the sample with $d_{Fe2}$=1.3 nm. The shape of this loop is
similar for all studied samples. In order to pin the magnetization
of the Fe1-layer by the bias CoO$_x$-layer the sample was cooled
down in a magnetic field of +4 kOe applied parallel to the sample
plane and measured at $T = 4$\,K. As a result in the field range
between -0.3 and -3.5 kOe the mutual orientation of two F-layers
is antiparallel because the magnetization of the Fe1-layer is kept
by the bias CoO$_x$-layer.
\begin{figure}[t]
\centering{\includegraphics[width=0.9\columnwidth,angle=-90,clip]{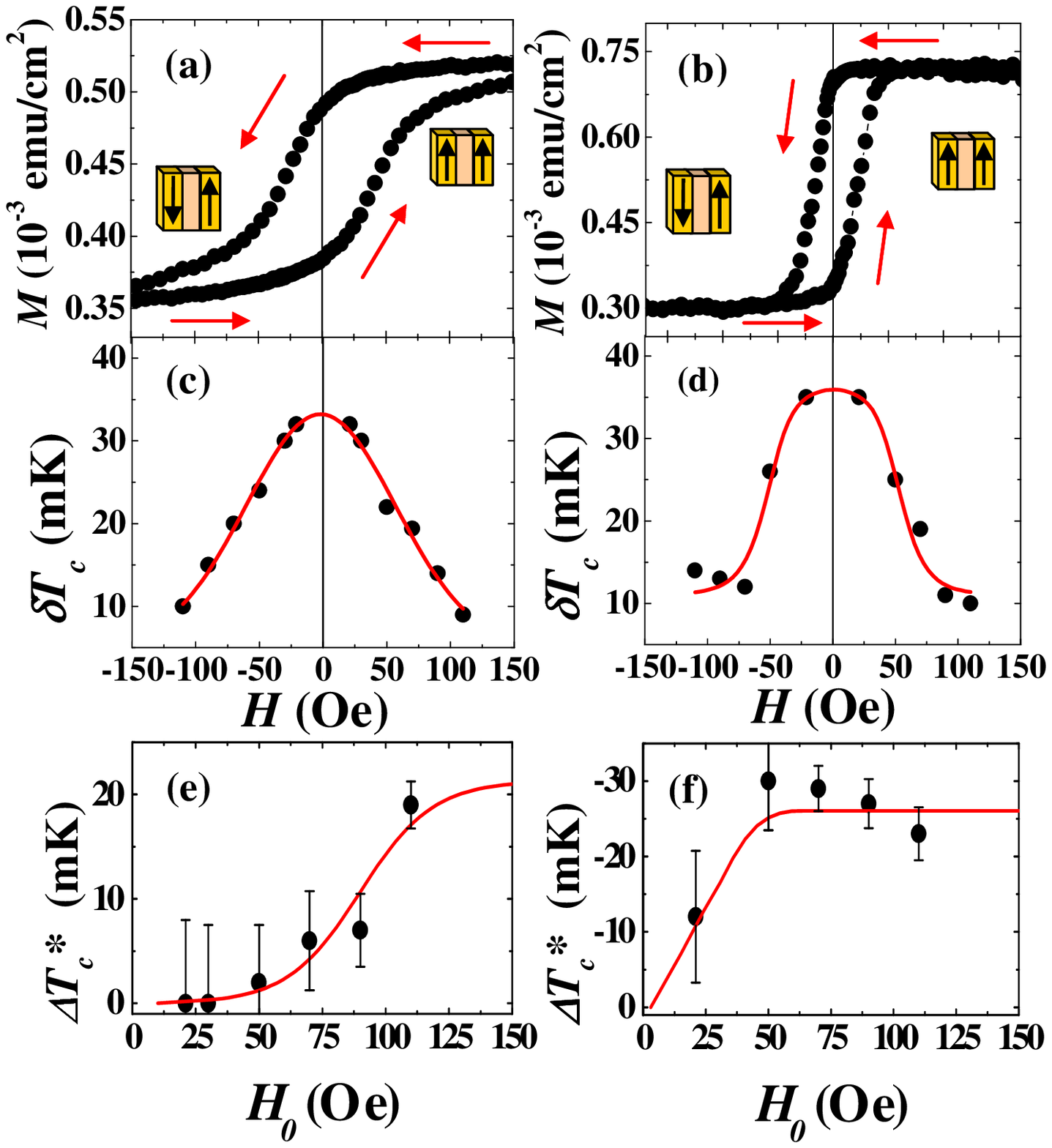}}
\caption{(Color online) The minor hysteresis loops for the spin
valve samples with $d_{Fe2}$=0.5 nm (a) and $d_{Fe2}=$1.3 nm (b)
and the dependencies of $\delta T_c$ and $\Delta T_c^*$ on the
magnetic field value: the panels (c) and (e) correspond to the
sample with $d_{Fe2}$=0.5 nm and the panels (d) and (f) - to the
sample with $d_{Fe2}$=1.3 nm.}
\end{figure}

A minor hysteresis loop was obtained by measuring the magnetization $M(H)$ with decreasing the field from +4 kOe down to -1 kOe and increasing the
field again up to +1 kOe. These minor loops for two representative samples with $d_{Fe2}$=0.5 and 1.3 nm are shown in Figs. 2a,b. As expected, the
magnitude of the change of $M$ is proportional to the thickness of the free F2-layer. The coercive field $H_c$, which is of the order of 30 Oe for
samples with $d_{Fe2}<$1\,nm, appreciably decreases down to $H_c \leq$20\,Oe with increasing the $d_{Fe2}$ above 1 nm. This clearly shows that the
domain state in the F-layers is progressively confined to smaller fields with increasing the $d_{Fe2}$. Usually in thin films the N´eel type domain
structure is realized. A decrease of $H_{\rm c}$ with increasing the Fe2-layer thickness takes place due to a decrease of the pinning of the domain
walls by interfaces arising due to the roughness and surface anisotropy.

In order to study the influence of the mutual  orientation of the magnetizations on $T_c$ we have recorded the change of the resistivity upon
transition to the SC state at the two in-plane switching field values of $+H_0$ and $-H_0$. First, the samples were cooled down from room temperature
to $T$\,=\,4\,K at a magnetic field of 4\,kOe applied along the easy axis of the sample just as we did it when performing the magnetization
measurements. For this field both F-layers' magnetizations are aligned. Next, the $T$-dependence of the resistivity $R(T)$ was recorded  at 7 fixed
values of the in-plane field $\pm H_0$ in the range $|H_0|= 0 \div$ 110\,Oe (examples see in  Fig. 3).
\begin{figure}[t]
\centering{\includegraphics[width=0.7\columnwidth,angle=-90,clip]{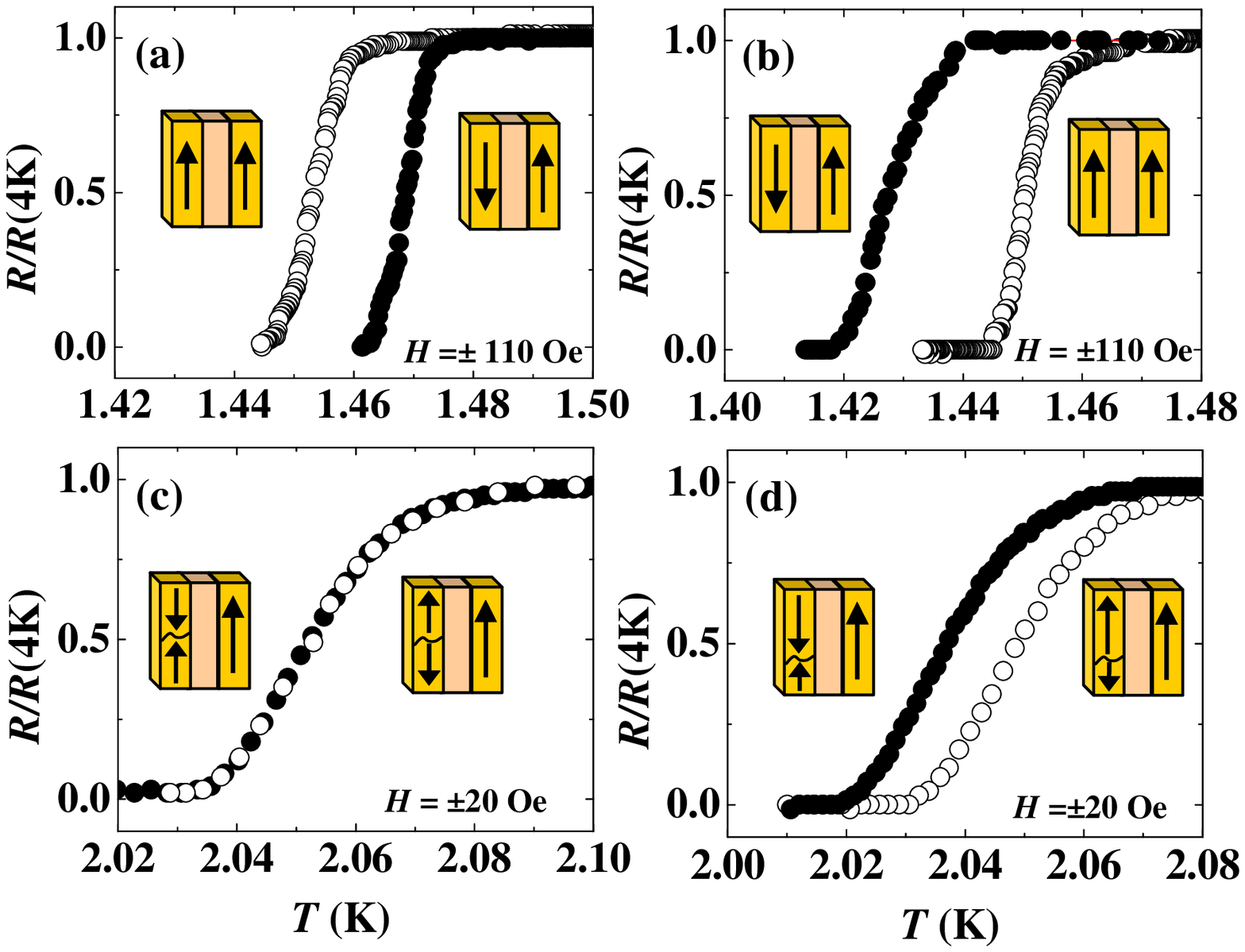}}
\caption{(Color online) The resistive superconducting transition
curves for the samples with $d_{Fe2}$=0.5 nm (a - $H$=110 Oe, c -
$H$=20 Oe) and 1.3 nm (b - $H$=110 Oe, d - $H$=20 Oe).}
\end{figure}
The dependencies of the width of the SC transition $\delta T_c$ and the magnitude of the spin valve effect $\Delta T_c^*=T_c(-H_0)-T_c(H_0)$ on $H_0$
are shown in Figs.~2c-2d and 2e-2f, respectively,  for two samples with $d_{Fe2}$=0.5\,nm and 1.3\,nm where they can be compared with the respective
minor hysteresis loops (Fig.2a, b). Both quantities have a clear correlation with the occurrence of the domain state: The presence of magnetic
domains broadens the SC transition and the spin valve effect can only be observed when domains {\it are suppressed} by magnetic field, i.e. when the
AP and P states are achieved. In this limit $\Delta T_c^*$ becomes equal to $\Delta T_c$. Remarkably, the sign of $\Delta T_c$ is positive for
$d_{Fe2}<$1\,nm and is negative for $d_{Fe2}\geq $1\,nm.

The detailed dependence of the spin valve effect  $\Delta T_c$ on
the thickness of the F2-layer is shown in Fig.~4. $\Delta T_c$
first increases with increasing $d_{Fe2}$ and has a sharp maximum
of 19 \,mK at $d_{Fe2}$=\,0.5\,nm. For larger thicknesses $\Delta
T_c$ decreases down to a very small value of 4\,mK for
$d_{Fe2}$\,=\,0.8\,nm suggesting a complete vanishing of the spin
valve effect at even larger thicknesses. Surprisingly, further
increasing the $d_{Fe2}$ in the interval $1\ {\rm nm}\leq d_{Fe2}
\leq$ 2.6 nm reveals the recovery of the effect but with the
negative sign.
\begin{figure}[t]
\centering{\includegraphics[width=0.6\columnwidth,angle=-90,clip]{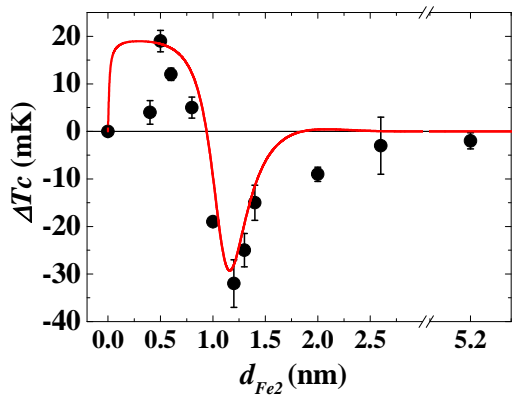}} \caption{(Color online) The dependence of the $T_c$ shift $\Delta
T_c=T_c^{AP} - T_c^{P}$ on the Fe2-layer thickness $d_{Fe2}$. The applied switching field $H = \pm 110$ Oe. Theoretical curve (solid line)
corresponds to the calculated function [W(0)-W(pi)]/W(0) (see Fominov et al. \cite{Fominov2}) normalized to our experimental data. This fit
gives $\xi_m=$0.9 nm, $l_m=$1.5 nm and a rough estimate of the quantum mechanical transparency of the S/F interface for the electrons
$T=$0.7 (see the text)}.
\end{figure}
$\Delta T_c(d_{Fe2})$ reaches its maximum negative value  of -33\,mK for $d_{Fe2}=$\,1.2\,nm and above this thickness smoothly approaches zero. Thus,
the $\Delta T_c(d_{Fe2})$-dependence exhibits a remarkable oscillating behavior in the thickness range 0.5 nm $\leq d_{Fe2} \leq$ 2.6 nm.

In the following we will discuss this striking observation  with regard to three scenarios: (i) occurrence of magnetic domains in the F-layers;
(ii) - spin accumulation in the S-layer; (iii) quantum interference of the Cooper pair wave function in the S/F multilayer. The theory of the spin
valve effect based on the S/F proximity effect \cite{Oh} predicts only the direct effect, i. e. $\Delta T_c >0$. The inverse effect with
$T_c^{AP}<T_c^P$ has been reported earlier for various systems \cite{Rusanov,Steiner,Stamopoulos,Zhu,Leksin1,Pena,Rusanov1,Singh}. Its origin was
discussed in terms of two, in fact conflicting, scenarios (i) and (ii). In model (i) magnetic domains may influence superconductivity in two
different ways. Rusanov et al. \cite{Rusanov} who studied Nb/Permalloy bilayers showed that the F-layer forms a domain state near its coercive field
and the S-layer experiences a lowered average exchange field seen by the Cooper pairs. This yields a direct effect which may be called a N\'{e}el's
domain wall induced enhancement of $T_c$. Another mechanism also involving domains invokes an interaction between outer F-layers in the F/S/F
trilayer, e.g., due to stray-field-induced magnetoelastic coupling \cite{Steiner,Stamopoulos,Zhu,Leksin1}. It yields the inverse effect since the
suppression of domains in the fully polarized state with the parallel alignment of the magnetizations of the F-layers drastically reduces the domain
induced stray fields normal to the layers which otherwise act on the S-layer and suppress the superconductivity.

Model (ii) is based on the giant magnetoresistance effect  and predicts an enhanced spin-dependent reflection of the spin-polarized charge carriers
at the S/F interface in the AP state. Hence, they can accumulate in the S-layer which gives rise to a reduction of the superconducting energy gap,
provided that the thickness of the S-layer is smaller than the spin diffusion length \cite{Pena,Rusanov1,Singh}.

For the interpretation of the negative sign of $\Delta T_c$  for thicknesses $d_{Fe2}\geq$\,1\,nm scenario (i) should be considered seriously since
even if the exchange coupling between the F-layers is negligible due to a thick enough 4\,nm Cu-interlayer \cite{Johnson}, the purely magnetostatic
reason for domain formation could play a role. A magnetic stray field normal to the In layer, which is a type I superconductor, might lead then to a
direct suppression of superconductivity. Indeed, the presence of the domain structure may be responsible for the broadening of the SC transition
$\delta T_c$ at small fields due to inhomogeneous magnetic field produced by the domain walls (Fig.~2c). However, for the values of the switching
field $H = \pm 110$\,Oe used to determine the $\Delta T_c(d_{Fe2})$-dependence in Fig.~4 the magnetizations of F1- and F2-layers are already
saturated in the P or AP configurations (Fig. 2b). The incomplete saturation where the domain induced stray fields normal to the S-layer may arise is
restricted to a small range of magnetic fields -50 Oe$<H_0<$50 Oe for the samples with $d_{Fe2}\geq$\,1\,nm, but just in this field range the inverse
spin valve effect is smallest compared to the values measured outside this range (Fig.~2f). Therefore we conclude that the influence of the domain
structure on the observed spin valve effect is negligibly small in our samples.

Scenario (ii) based on the spin imbalance mechanism can be  surely
excluded in our case because the enhanced spin-dependent
reflection of spin-polarized carriers at the S/F interfaces can
occur in the AP state of the F/S/F trilayer but not in our F1/F2/S
structures where the S-layer is not sandwiched between the
F-layers.

As to scenario (iii), indeed in a ferromagnetic layer the Cooper pair acquires a non zero momentum due to the Zeeman splitting of electron levels
and thus its wave function should oscillate in space (see, e.g., \cite{Buzdin_rev}). If the F-layer is sufficiently thin, the wave function reflected
from the surface of the F-layer opposite to the S/F interface can interfere with the incoming one. Depending on the layer's thickness the
interference at the S/F interface may be constructive or destructive. This should apparently lead to the enhancement of $T_c$ or its decrease,
respectively, thus naturally explaining our main result (Fig.~4).

Interestingly, there is a recent theory developed by Fominov et al. \cite{Fominov2,Fominov1} where the same spin switch scheme F1/F2/S is considered.
The starting points there do not strictly comply with the properties of our samples: F-layers were assumed to be weak ferromagnets, simplified
boundary conditions were taken implying a 100 \% transparency of the F2/S and F1/F2 interfaces for the electrons and superconductivity in the "dirty"
limit ($l_m <\xi_m$) were assumed. Here $l_m$ is the mean free path of conduction electrons. In our samples the F-layer made of iron is a strong
ferromagnet with $\xi_m \sim$1\,nm. In this case the transparency of the S/F interface should be reduced due to the exchange splitting of the
conduction band in the F-layer \cite{Lazar}.  Also the "dirty" limit is not realized owing to a small value of $\xi_m$. Finally the considered model
does not involve the presence of the N-layer and assumes the F1-layer to be a half infinite ferromagnetic layer.

However, it is known that in practice the S/F proximity theories developed for the "dirty" limit deliver reliable results even beyond the domain of
their applicability. Indeed, despite these differences we  were able to obtain a reasonably good qualitative agreement between this theory and our
experimental results as demonstrated by the fit curve to the experimental $\Delta T_c(d_{Fe2})$-dependence in Fig.~4. An appreciable discrepancy with
the experimental data point $d_{Fe2}$=0.4\,nm occurs most probably because at this thickness a transition from a continuous to an island like Fe film
at even smaller thicknesses $d_{Fe2}$ does take place. The fit parameters turned out to be quite realistic. We obtained  $\xi_m=$0.9\,nm and
$l_m$=1.5\,nm, confirming that our samples satisfy the "clean" limit ($l_m >\xi_m$), and the quantum mechanical transparency of the S/F interface for
the electrons $T$\,=\,0.7. The reasonable values of the fit parameters and the fact that the theory correctly describes the observed oscillation of
the $T_c$-shift $\Delta T_c$ gives additional arguments favoring the S/F proximity effect as the origin of our striking observation, which however
stands on its own regardless a specific theoretical model.

In summary, we have presented experimental evidence for the oscillating behavior of the spin valve effect in a ferromagnetic/superconductor
multilayer F1/N/F2/S with a varied thickness of the ferromagnetic F2-layer. We have observed the direct spin valve effect for F2-layer thicknesses
smaller than the decay length $\xi_m$ of the Cooper pair wave function in the F2-layer and the inverse spin valve effect for larger thickness up to
$2.5\xi_m$. The analysis of the data suggests that the inverse spin valve effect is likely caused by the interference effects for the superconducting
pairing function reflected from both surfaces of the F2-layer.

\begin{acknowledgements}
The authors are grateful to A. F. Volkov, L. R. Tagirov  and Ya.
V. Fominov for useful discussions and R. Klingeler for the support
in magnetic measurements. The work was supported by the RFBR
(grants No. 08-02-00098, 08-02-91952-NNIO, 11-02-01063 and by the
DFG (grant BU 887/13-1).
\end{acknowledgements}

\end{document}